\documentclass[conference]{IEEEtran}
\IEEEoverridecommandlockouts
\usepackage{cite}
\usepackage{amsmath,amssymb,amsfonts}
\usepackage{algorithmic}
\usepackage{graphicx}
\usepackage{textcomp}
\usepackage{xcolor}
\usepackage{subfigure}
\def\BibTeX{{\rm B\kern-.05em{\sc i\kern-.025em b}\kern-.08em
    T\kern-.1667em\lower.7ex\hbox{E}\kern-.125emX}}
\begin{document}

\title{Impact of Network Delay and Decision Imperfections in IoT Assisted Cruise Ship Evacuation\\
{\footnotesize}
\thanks{This research was partially funded by the EU H2020 Program’s IoTAC Research and Innovation Action under Grant No. 952684 at the Institute of Theoretical and Applied Informatics, Polish Academy of Sciences (IITIS-PAN), partially by the National Natural Science Foundation of China (NSFC) under Grant No. 51979216 and the Natural Science Foundation of Hubei Province, China, under Grants No.2021CFA001 and 20221j0059.}
}

\author{\IEEEauthorblockN{Yuting Ma}
	\IEEEauthorblockA{IITiS-PAN, Gliwice, Poland\\
		\& Wuhan University of Technology\\
		Wuhan, China\\
		yma@iitis.pl}
	\and \IEEEauthorblockN{Erol Gelenbe}
\IEEEauthorblockA{IITiS-PAN, Gliwice, Poland\\
\& Universit\'{e} C\^{o}te d'Azur, Nice, France\\
\&Ya\c{s}ar University, Bornova, Izmir, Turkey\\
seg@iitis.pl}
\and
\IEEEauthorblockN{Kezhong Liu}
\IEEEauthorblockA{Wuhan University of Technology\\
Wuhan, China \\
kzliu@whut.edu.cn}
}

\maketitle

\begin{abstract}
Major challenges of assisting passengers to safely and quickly escape from ships when an emergency occurs, include complex realistic features such as human behavior uncertainty, dynamic human traversal times, and the computation and communication delays in the systems that offer advice to users during an emergency. In this paper, we present simulations that examine the influence of these key features on evacuation performance in terms of evacuation time. The approach is based on our previously proposed lookup table-based ship passenger evacuation method, i.e., ANT. The simulation results we present show that delays in the users' reception of instructions significantly impair the effectiveness of the evacuation service. In contrast, behavior uncertainty has a weaker influence on the performance of the navigation method. In addition, these effects also vary with the extent of the behavior uncertainty, the dynamics of the traversal time distributions, and the delay in receiving directions. These findings demonstrate the importance of carefully designing evacuation systems for passenger ships in a way that takes 
 into account all realistic features of the ship's indoor evacuation environment, including the crucial role of information technology. 
\end{abstract}


\section{Introduction}
One of the main safety requirements for passenger ships is offering a reliable and efficient emergency evacuation service to passengers. Therefore, evacuation analysis is of primary importance even in the first stages of the design of a vessel \cite{nasso2019simplified,wang2022numerical,arshad2022determinants}. The International Maritime Organization (IMO) has issued many Circulars about evacuation analysis and procedures for passenger and ro-ro ships. For example, in 2016, IMO approved a new version of the guidelines for evacuation analysis for new and existing passenger ships to enhance the ability to  evacuate passenger vessels in case of accidents \cite{international2016guidelines}. 

\begin{figure}[t]
\centering
\includegraphics[width=8.6cm,height=3cm]{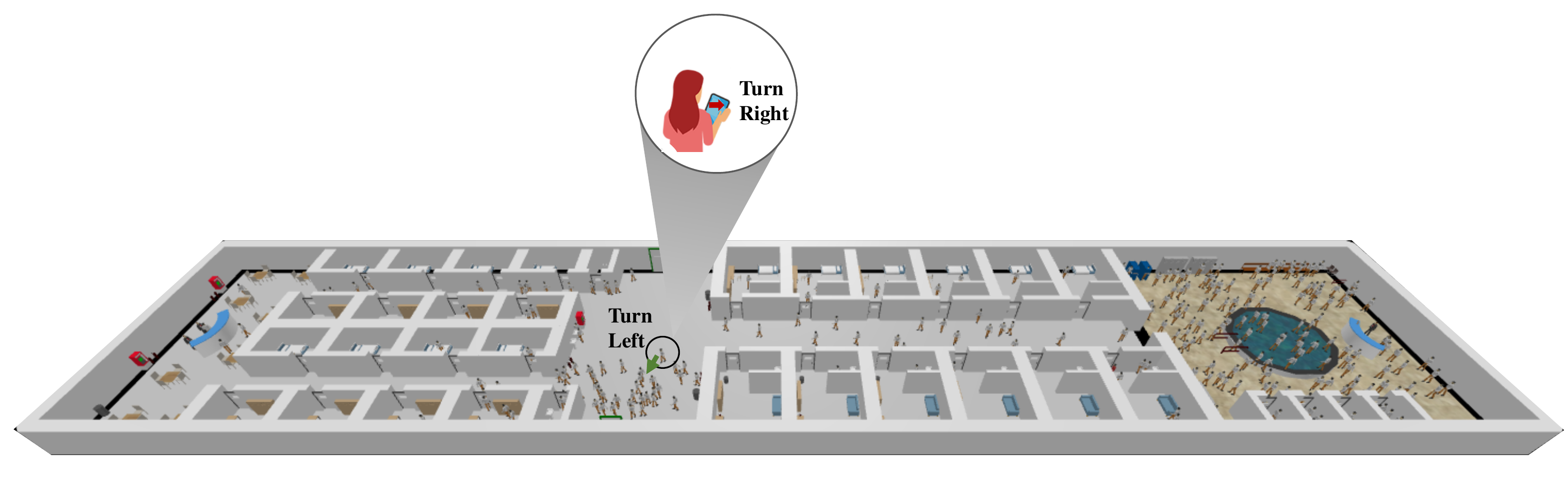}
\caption{Schematic representation of behavior uncertainty of a human user during an emergency.} 
  \label{Fig1} 
\end{figure}

However, much of the research in this area does not consider many realistic features which may have a huge impact on both user safety and navigation efficiency \cite{Huibo2,xie2022integrated,fang2022simulation,wang2023novel}, such as variations in traversal times across passages and staircases due to vessel motion conditions, human behavior uncertainty caused by missing or misunderstanding instructions due to panic, noise, and overcrowding as illustrated in Figure \ref{Fig1}, and delays in the arrival of correct instructions due to computation and communication delays
in heavily loaded communication networks that monitor and interact with the physical world. As a result, evacuation metrics such as evacuation time and passenger congestion calculated by these methods are not necessarily in line with those in actual situations. Thus the design of the evacuation system based on such analyses may not ensure the needed safe and timely navigation of evacuees.
Hence, this paper investigates the influence of such key features on the performance of the navigation method proposed in \cite{ma2020ant} under simulated emergency conditions.  

In this paper, we make some model simplifications and assume that the inclination angle of the damaged ship changes at regular intervals, so that the traversal times encountered across each individual corridor and staircase accordingly change due to the variation of the inclination status. In addition, some significant extensions are made in this paper with respect to previous work \cite{ma2020ant} by introducing an error probability $\textit{PoE}$. Then we change the delay probability $\textit{PoD}$ and its extent $\textit{SoD}$,  to learn the variation in evacuation performance. In addition, we explore the combined effects of both behavior uncertainty and service delay on evacuation times.

\section{Related work}
$\textbf{Global}$ $\textbf{path-planning}$ algorithms (e.g., Dijkstra, A*, RRT, Ant Colony Optimization, Genetic algorithm, etc.,) have often been successfully used in evacuation routing to minimize
path length or to maximize the distance from the hazard nodes to the exit node, based on 2D/3D maps of a built environment \cite{Desmet, Kokuti2}. However, the resulting navigation paths provided by those methods are not necessarily useful since evacuees are likely to encounter unexpected hazards due to 
changing emergency conditions, such as the spread of a fire, flooding of parts of the ship, or cascaded failures in emergency management systems themselves. Some work has attempted to address the dynamics of danger through the Expected Number of Oscillations (ENO) concept \cite{wang2014oscillation}, that quantifies the dynamics of the emergency and explores navigation paths that have the smallest probability of changing repeatedly.

Furthermore, such global path planning methods require that the system be completely
specified regarding the parameters needed to choose optimal paths, in contradiction with realistic situations where emergency guidance is needed. Therefore, follow-up studies relax the requirement that all the information about the hazards and exits is precisely known prior to path computation, which enhances their applicability to address practical scenarios \cite{Filip,Huibo2}. 

Using $\textbf{local}$ $\textbf{path-planning}$ methods such as Artificial Potential Fields \cite{Schmajuk}, and the
 best direction for users based on conditions in their local neighborhood \cite{tseng2006wireless, Kokuti}
together with techniques such as partial reversal, users can avoid entering hazardous areas. However, these methods neglect the fact that the evacuation system for passenger ships must offer safety guarantees, and that the emergency navigation system must offer directions that guarantee a time to the exit that does not exceed a specified upper bound (i.e., the survival or capsizing time) even under worst-case circumstances (e.g., the case where the roll angle reaches the 30$^{\circ}$) in the presence of dynamic hazards and ship motion. 

However our earlier work \cite{ma2020ant} proposed an emergency navigation system based on condition which neglect the effect of uncertainty regarding the evacuees' behavior due to panic or errors, and the possible lack of up-to-date information regarding  estimates of typical delays due to computtaional or communication overload in the information processing system. Thus in the present paper we specifically address the effect of delays in providing accurate guidance to end users, as well as the possibility that human evacuees may make mistakes in applying the guidance instructions due to panic or lack of attention. 

\section{Impact of delay in user navigation}
The lookup tables in a static scenario guarantee that users at different locations can identify the hazard-free direction along which they can arrive at the exit with minimum delay,  while not exceeding the worst-case time bound from the current location to the exit. However, walking conditions are dynamic in practice
due to individual differences of the passengers, as well as other factors such as lighting, possible smoke conditions, panic, and the ship's motion.  These lookup tables then need to be recomputed due to the
overall system dynamics over time. Once there is a significant change in the tuples contained in the lookup table at each node in the navigation graph, the corresponding navigation direction needs to be updated for each of the users. However, wireless sensor network (WSN, and other network and computational server congestion effects may occur, so that updates of the navigation direction can be delayed, or even occur with errors due to the use of incomplete data in the decision algorithms \cite{shi2022wireless,dong2022faster,ma2022multi}. Thus we analyze and discuss in the following section the impact of these dynamics on the effective delays that may be experienced by the users.

\subsection{Simulation setup}
\begin{figure*}[ht]
\centering
\subfigure[]{
\includegraphics[width=8cm]{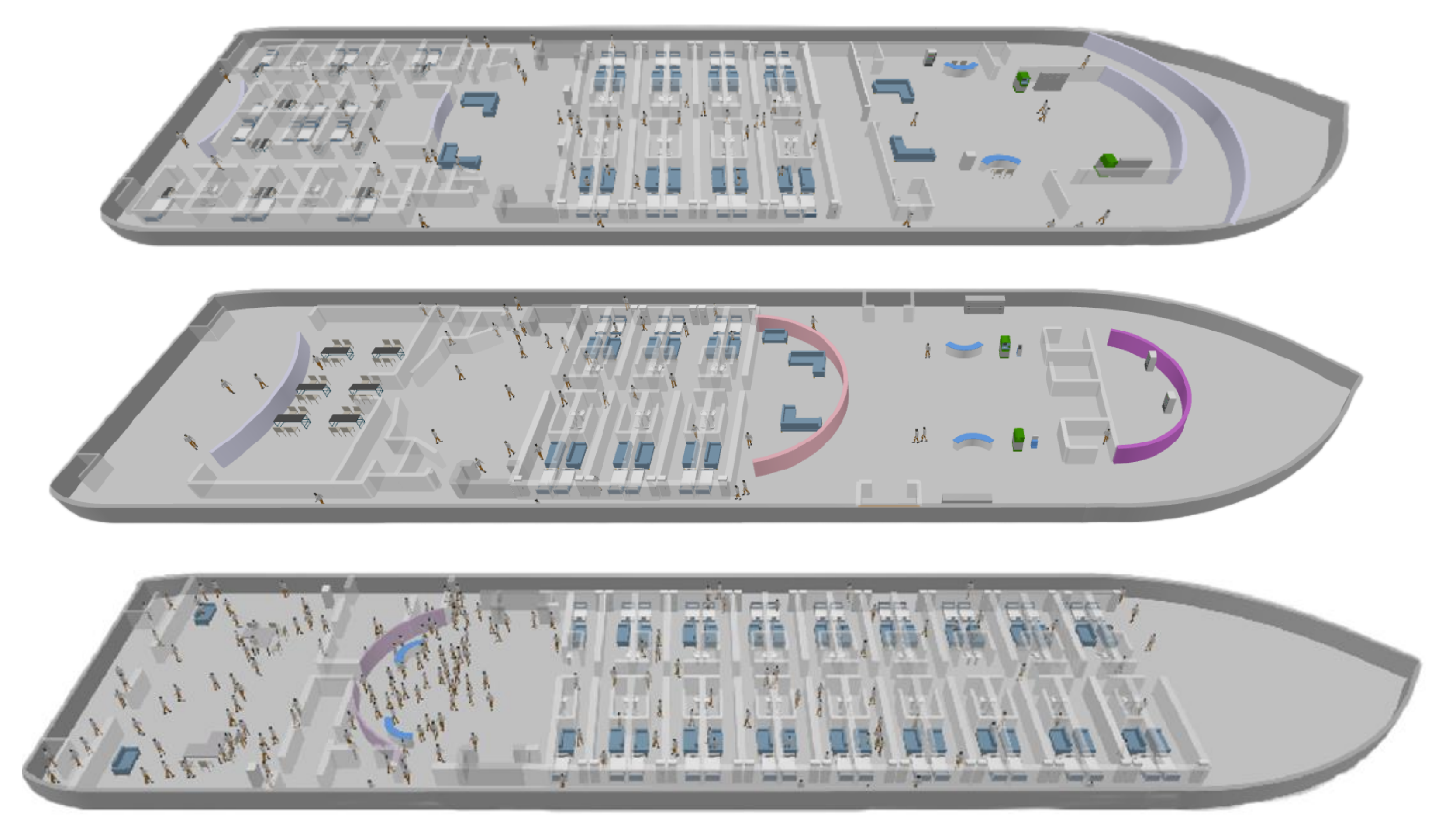}
}
\subfigure[]{
\includegraphics[width=8cm,height=4cm]{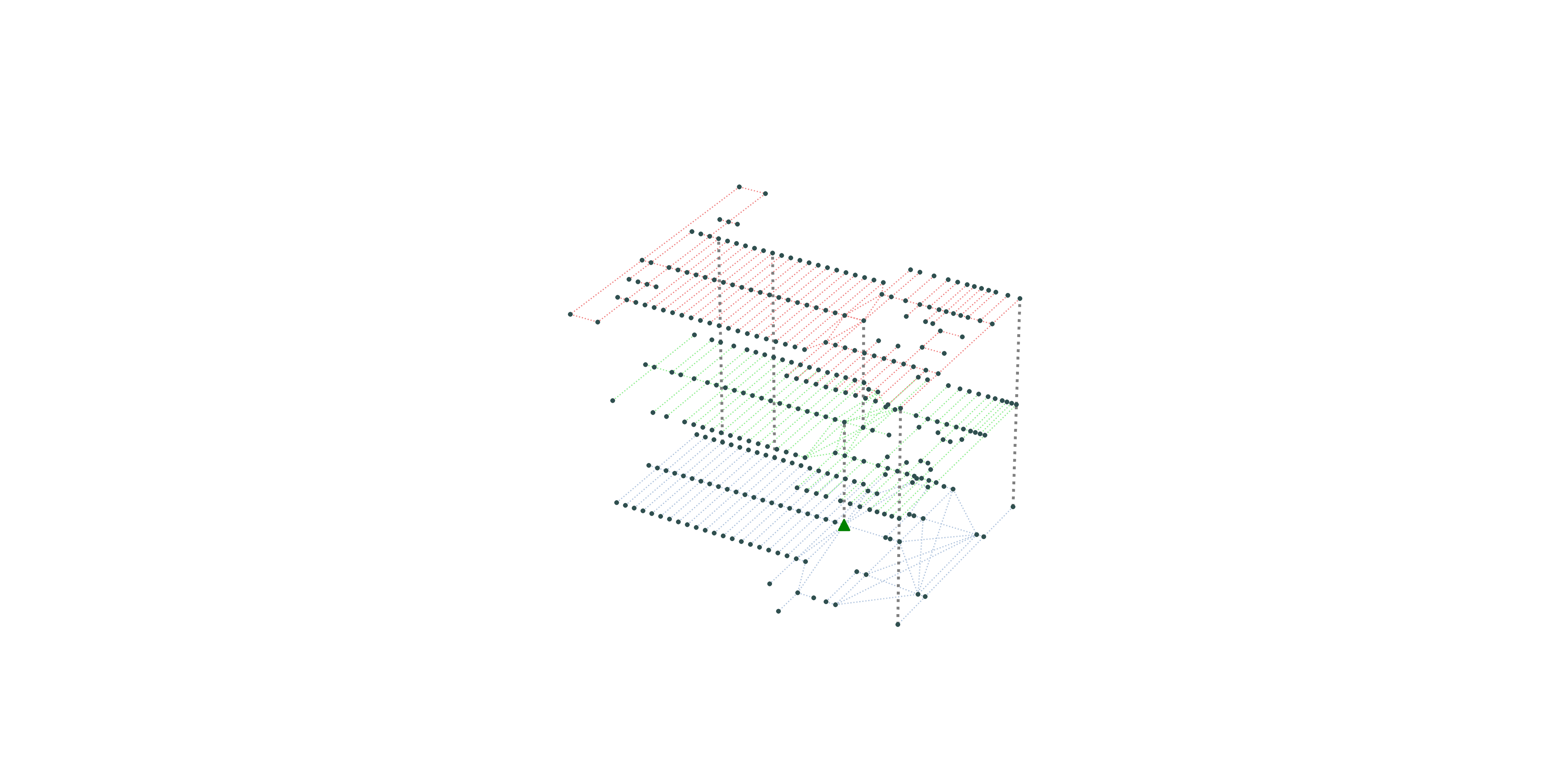}
}
\caption{The second, third, and fourth floors of the Yangtze Gold 7 Cruise.} 
  \label{Fig2} 
\end{figure*}

The simulated indoor environment is the second, third, and fourth floors of the Yangtze Gold 7 Cruise, as shown in  Figure \ref{Fig2}, where 346 navigation nodes are deployed in the simulated floors with one exit node. The number of passageway segments and staircase segments is 600 and 5, respectively. In our simulation settings, the worst-case traversal time $\textit{T}_{\rm {W}}$($\overrightarrow{v_{i}v_{j}}$) across each segment $\overrightarrow{v_{i}v_{j}}$ is calculated according to the worst-case traversal speed, which is set to 0.067 m/s, and the typical traversal time experienced in traversing each segment is set to a random value in the interval $\left[\textit{T}_{\rm {N}}(\overrightarrow{v_{i}v_{j}}), \textit{T}_{\rm {W}}(\overrightarrow{v_{i}v_{j}}\right]$. $\textit{T}_{\rm {N}}$($\overrightarrow{v_{i}v_{j}}$) = 0.67m/s, which is set according to the average movement speed of passengers on the passenger ships where the walking condition remains horizontal. In addition, the deadline $\textit{T}_{\rm{D}}$ for ship capsizing is calculated as follows:
\begin{equation}
    \textit{T}_{\rm{D}} = \textit{T}_{\rm{S}} - \textit{T}_{\rm{A}} - \textit{T}_{\rm{EL}}.
\end{equation}
where $\textit{T}_{\rm{S}}$ denotes the total survival time until the ship will capsize, and $\textit{T}_{\rm{A}}$ is the awareness time beginning upon initial notification of an emergency and ending when passengers accept the situation and start to move based on the provided navigation direction. $\textit{T}_{\rm{EL}}$ denotes the sum of embarkation time and the launching time, i.e., the time required to provide for abandonment by the total number of persons on board. According to the guidelines approved by the Maritime Safety Committee (MSC), $\textit{T}_{S}$ equals to 60 minutes, $\textit{T}_{\rm{A}}$ equals to 5 minutes, and  $\textit{T}_{\rm{EL}}$ equals to 25 minutes in our simulations. That is, $\textit{T}_{\rm{D}}$ equals to 30 minutes unless stated otherwise. In addition, the change interval of typical traversal time across each segment is set to 5s.  

\subsection{Simulation results}

We first examine the manner in which performance is affected by the delay in the computing and communication system which advises the users, and combine a probabilistic and temporal representation of the imperfection with which the system provides advice.

We assume that a delay in the communication and computing system can occur with some probability $\textit{PoD}$ at any of the user nodes. Data about the system as a whole is assumed to be updated in unit time steps, and we define $\textit{SoD}$ as being the number of unit time steps regarding the delay with which the
data is used. Thus, for instance, $\textit{SoD}=5$ means that at some time $\textit{t}$ the computational system will provide
decisions for the users based on the data that was available at time $\textit{t}-5$.

\begin{figure}[ht]
  \centering 
    \includegraphics[width=3.4in,height=6cm]{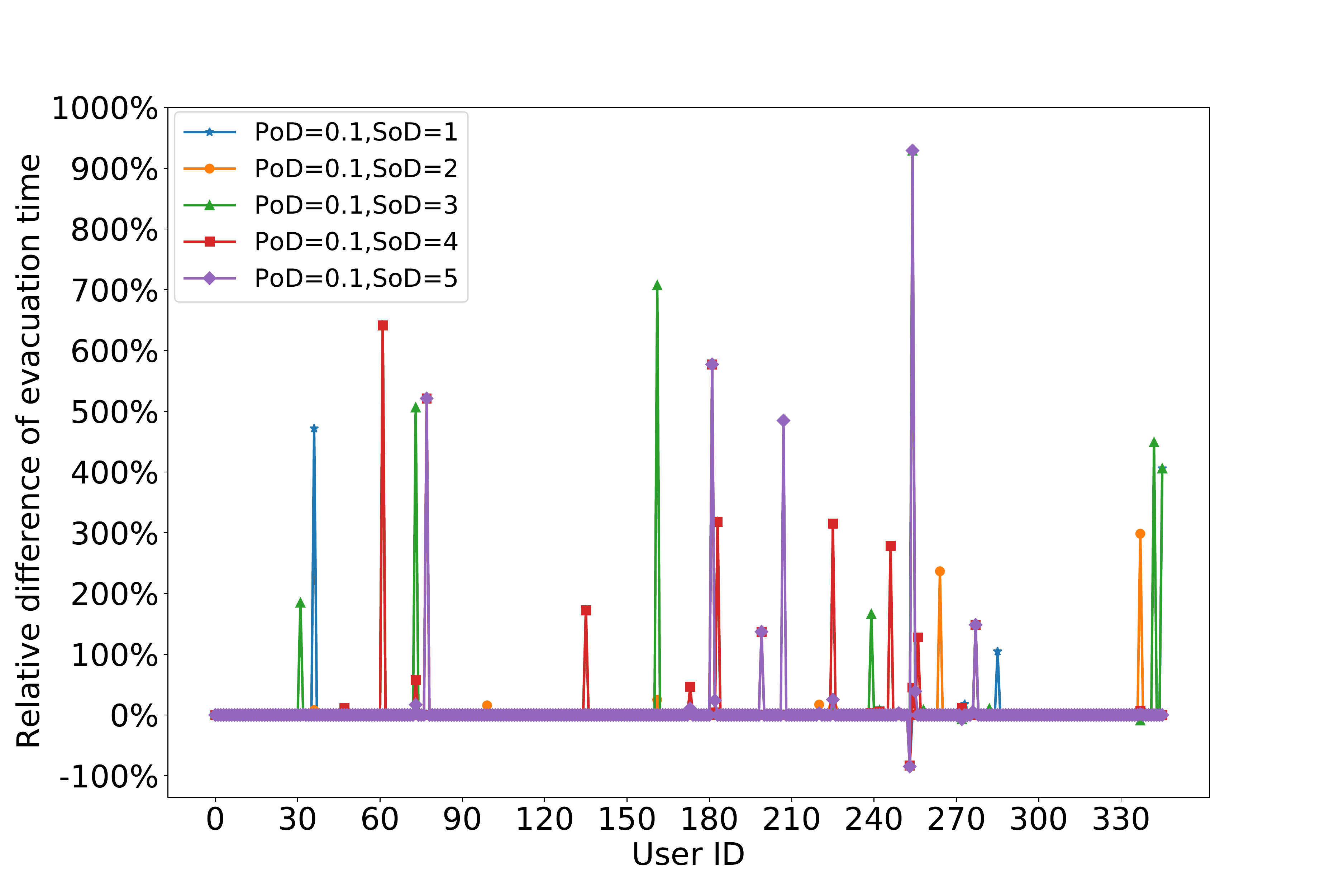} 
  \caption{Relative difference in evacuation time between guiding without delay and guiding with different extents of delay.} 
  \label{Fig3} 
\end{figure}

Thus in Figure \ref{Fig3} we consider results for the following parameters $\textit{PoD}=0.1$ and $\textit{SoD}=1,~2,~3,~4,~5$. More particularly, we show the relative difference in 
 evacuation time for each user with these five different delay values for all of the unique Node identifiers $0$ to $345$ (assigned when the navigation network is constructed) in order to visualize the simulation results. The relative difference in evacuation times for each node is defined as follows:
\begin{equation}
\Delta(i, PoD, SoD)=\frac{T(i, PoD, SoD)-T(i)}{T(i)},
\end{equation}
where $T(i)$ denotes the ``ideal'' evacuation time for a user at location (node) $i$ all the way to the exit, when there is no delay, i.e. $PoD=0$, while $T(i, PoD, SoD)$ is the corresponding value with probability $\textit{PoD}$ that there will be a delay of value $\textit{SoD}$. We clearly see that guiding evacuees with up-to-date navigation suggestions has better performance with respect to the evacuation time, and therefore the observed relative differences are positive in most cases. But it is worth noticing that there are still a few cases where the performance of guiding with up-to-date navigation directions is even worse than that of guiding with outdated navigation suggestions. It is mainly because the navigation based on up-to-date suggestions is not the optimal scheme for addressing the dynamics of walking conditions. 

\begin{figure}[ht]
  \centering 
    \includegraphics[width=3.4in,height=6cm]{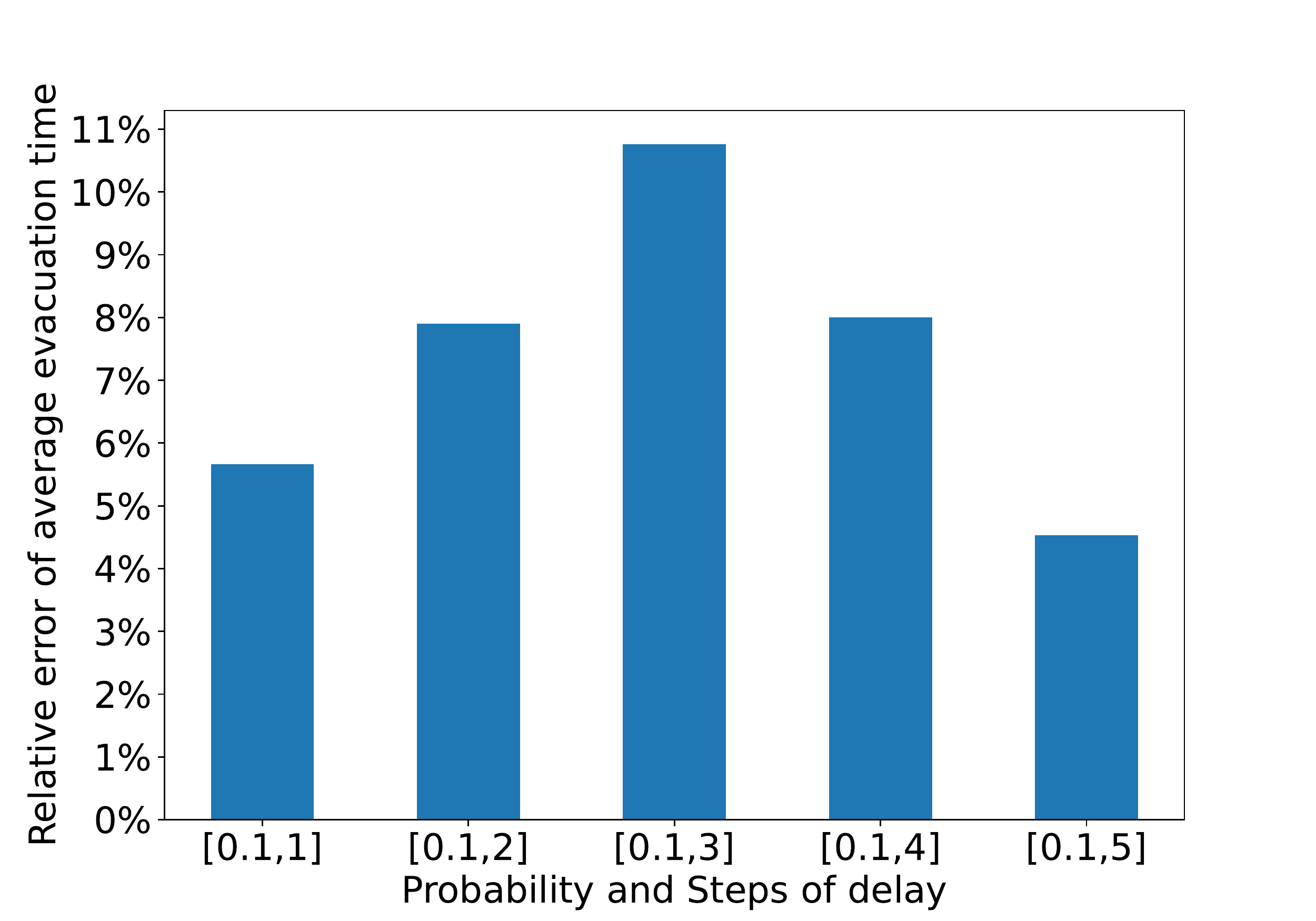} 
  \caption{Relative difference in average evacuation time between guiding without delay and guiding with different extents of delay.} 
  \label{Fig4} 
\end{figure}

In addition, we also simulate the average evacuation time with different delay values for all the users located in the 346 different user nodes. Figure \ref{Fig4} plots the relative difference in average evacuation time defined as:
\begin{eqnarray}
&&\Delta_{AVG}(PoD,SoD)\nonumber\\
&&~~=\frac{\sum_{i=0}^{345}T(i,PoD,SoD)-\sum_{i=0}^{345}T(i)}{\sum_{i=0}^{345}T(i)}.
\end{eqnarray}
It is observed that the relative difference in average evacuation time initially increassd with the delay  $\textit{SoD}$, but when $\textit{SoD}$ exceeds $3$, the relative difference does not continue to increase. Thus it appears that for the parameters we have chosen there is a criticality point for the worst evacuation performance, and in our simulation, it is at $\textit{SoD} = 3$.

\begin{figure}[ht]
\centering
\subfigure[Average evacuation time of the 150 users]{
\includegraphics[width=3.33in,height=6cm]{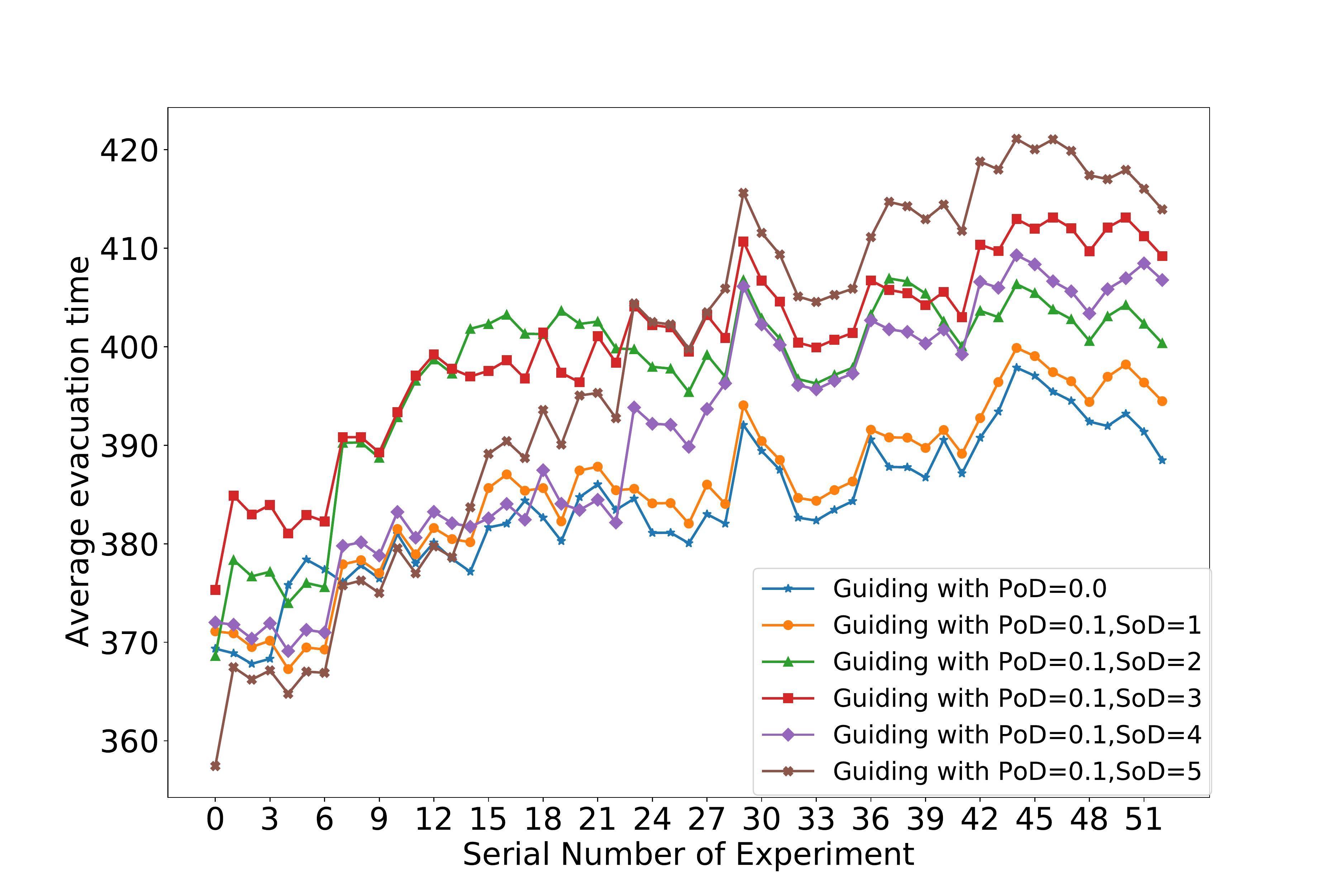}
}
\subfigure[Relative difference in the average evacuation time of the 150 users]{
\includegraphics[width=3.33in,height=6cm]{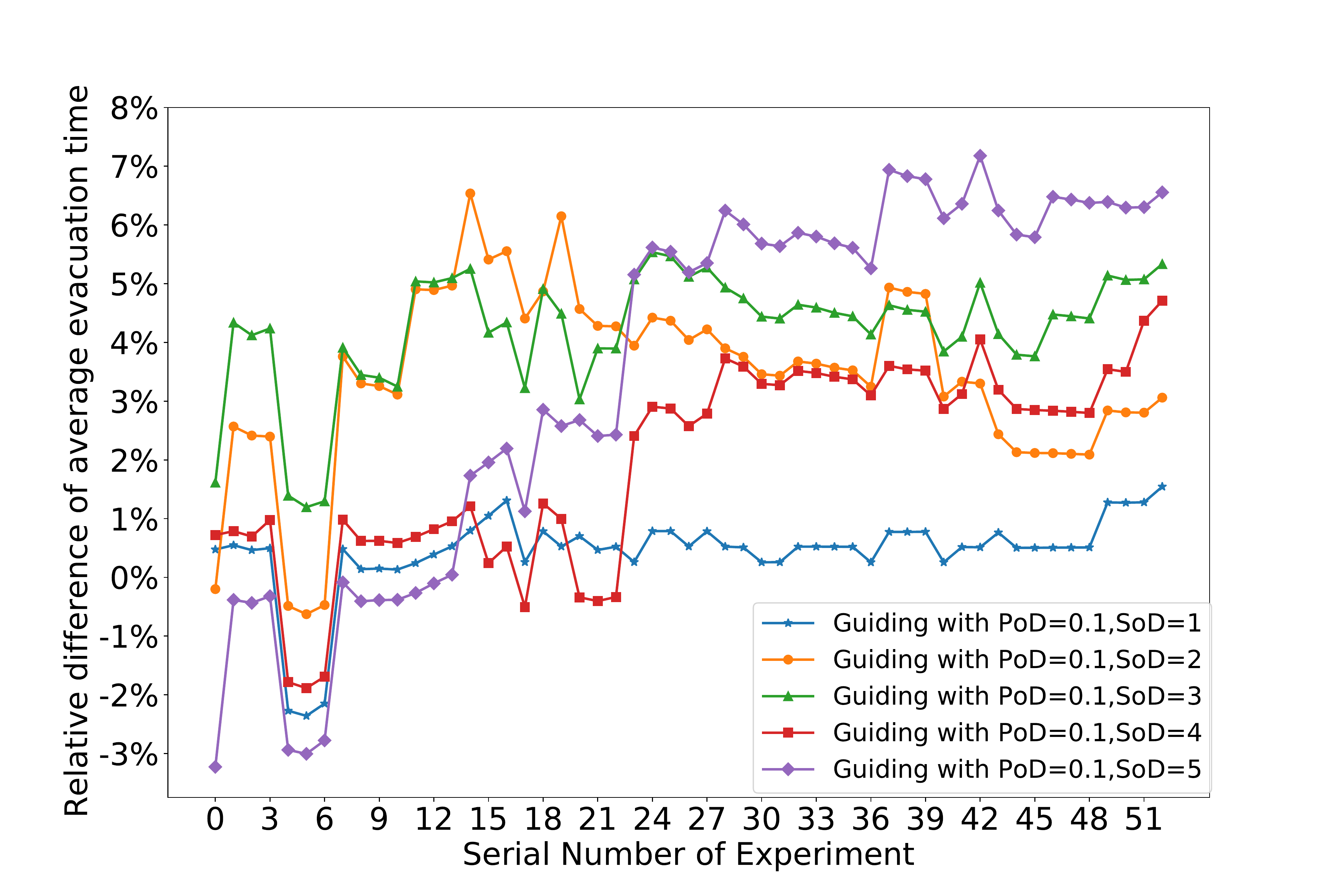}
}
\caption{Average evacuation time and the relative difference of the 150 users.} 
  \label{Fig5} 
\end{figure}

We also carry out a group of simulations, where $150$ users are randomly deployed at the user nodes, and Figure \ref{Fig5} shows the average evacuation time and the relative difference of the $150$ users from $53$ runs of the simulation. The observation from Figure \ref{Fig5} aligns with the findings presented in Figures \ref{Fig3} and \ref{Fig4}.             

\begin{figure*}[ht]
\centering
\subfigure[SoD = 1]{
\includegraphics[width=5.71cm]{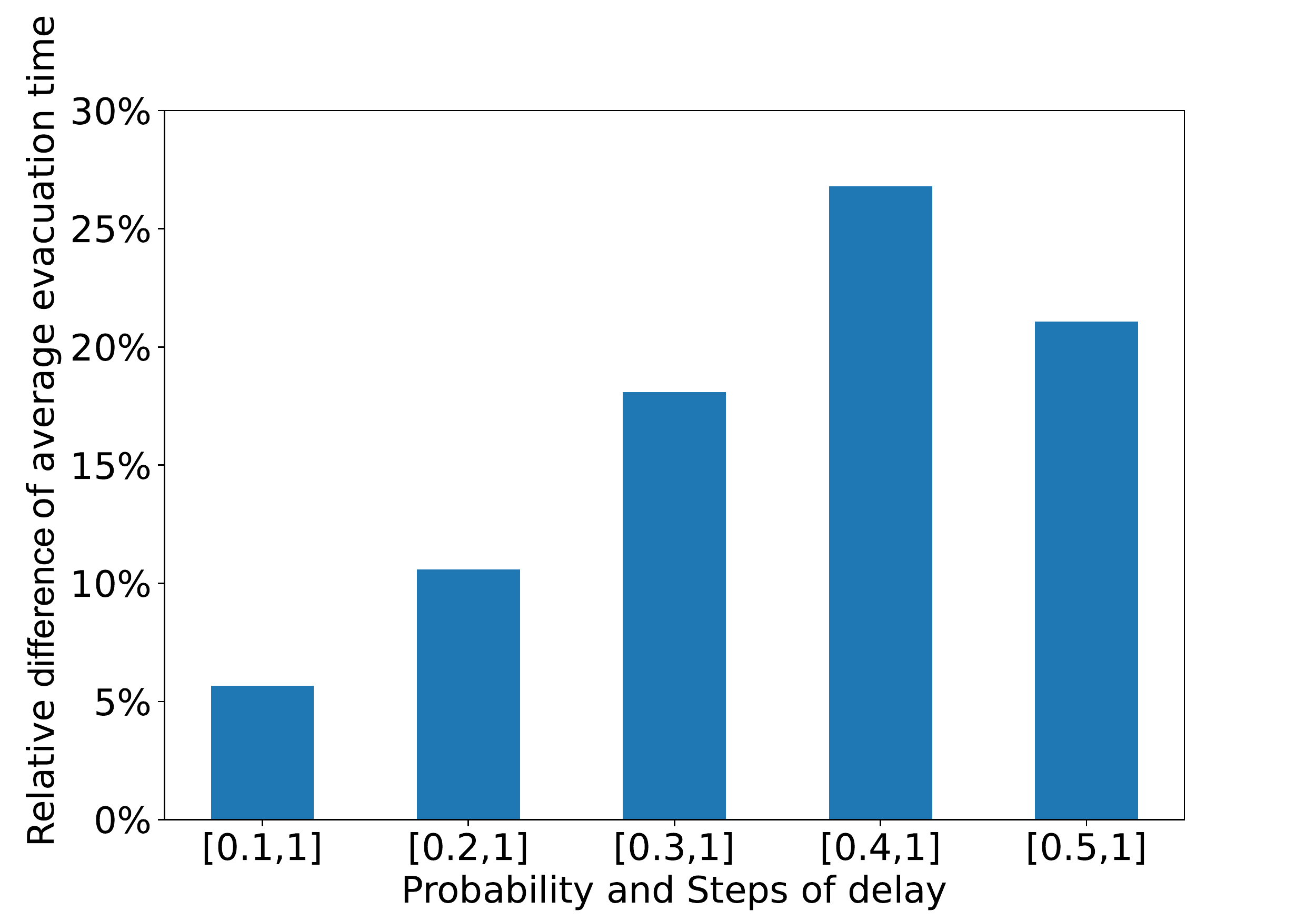}
}
\subfigure[SoD = 2]{
\includegraphics[width=5.71cm]{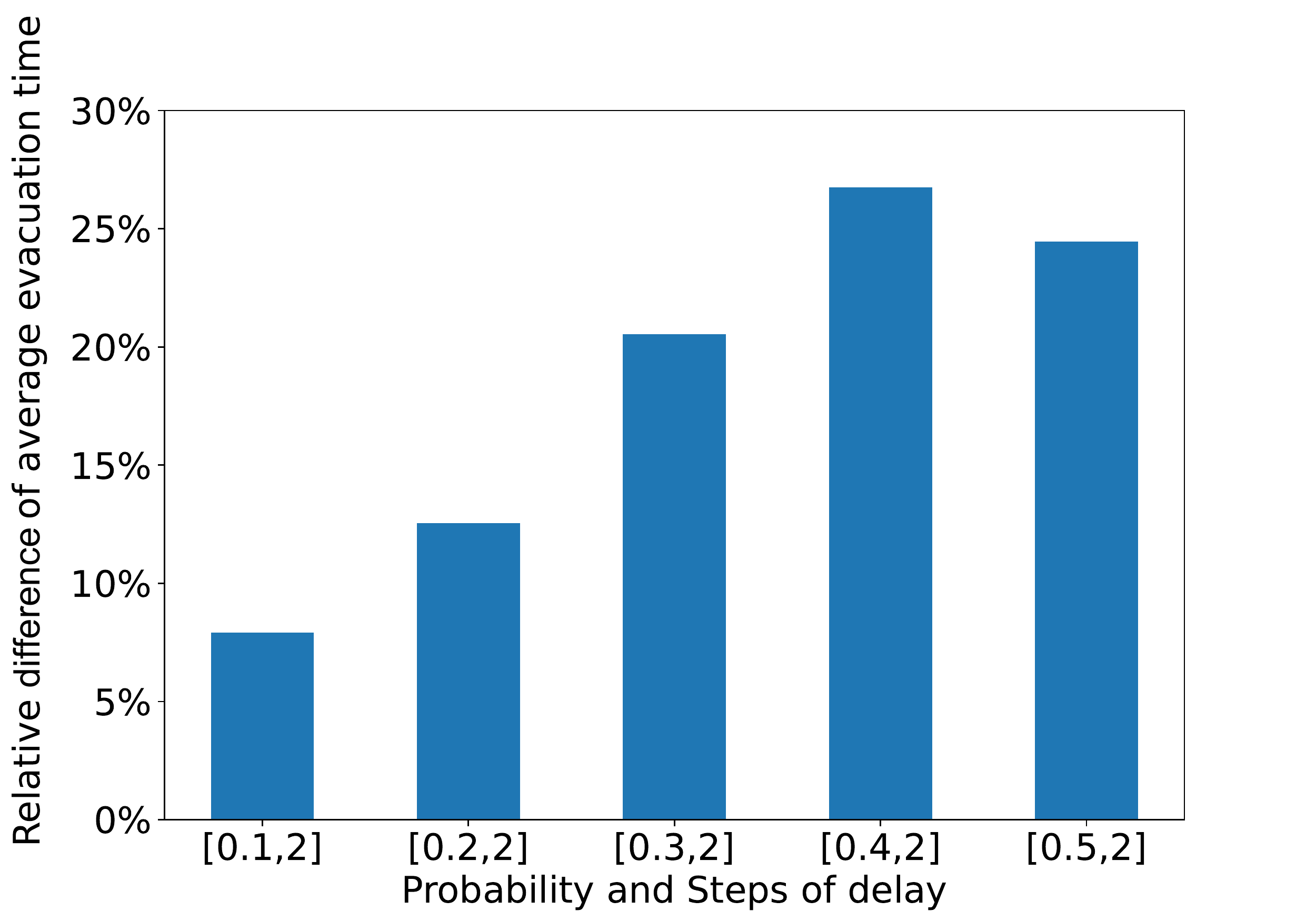}
}
\subfigure[SoD = 3]{
\includegraphics[width=5.71cm]{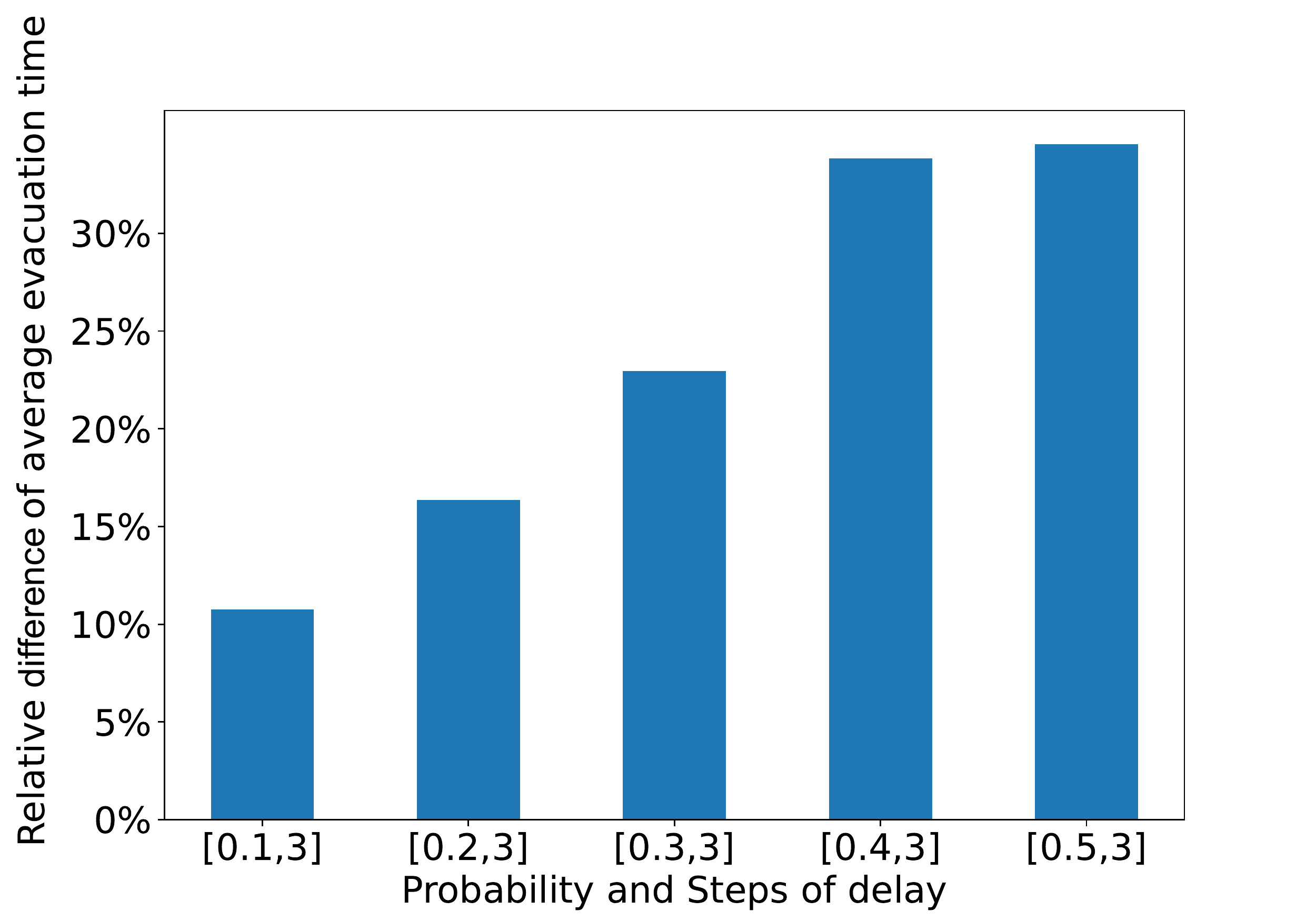}
}
\caption{Relative difference of average evacuation time of the 346 users for PoD = 0.0, 0.1, 0.2, 0.3, 0.4, 0.5 and SoD=1, 2, and 3, respectively.} 
  \label{Fig6} 
\end{figure*}

We further evaluate the influence of the different possibilities of delay in navigation service on evacuation time. Specifically, we conduct a group of simulations, where the possibility of delay in navigation service at each node is set to $0.0,~0.1,~0.2,~0.3,~0.4$, and $0.5$. The delay value $SoD$ is set to $1,~2$, and $3$. Figure \ref{Fig6} shows the relative difference in average evacuation time for 346 users at different user nodes. We can see that the relative difference in average evacuation time is increased with the possibility of delay in navigation service and achieves its maximum when $PoD = 0.4$ for $SoD = 1,~2,~3$.  

\section{Impact of behavior uncertainty}
When an emergency occurs, the evacuees may move due to panic or difficulties in reading instructions, along a random direction instead of the direction provided by the evacuation system. In order to explore the resulting performance, we represent this effect by an error probability and measure the evacuation time accordingly. 346 users are inserted into the simulated network in Section III. In addition, in this group of simulations, the typical traversal time across each segment is static in order to focus on the impact of the behavior uncertainty. The simulation parameters used in this section are the same as those in Section III.      

\begin{figure}[ht]
  \centering 
  \includegraphics[width=3.4in,height=10cm]{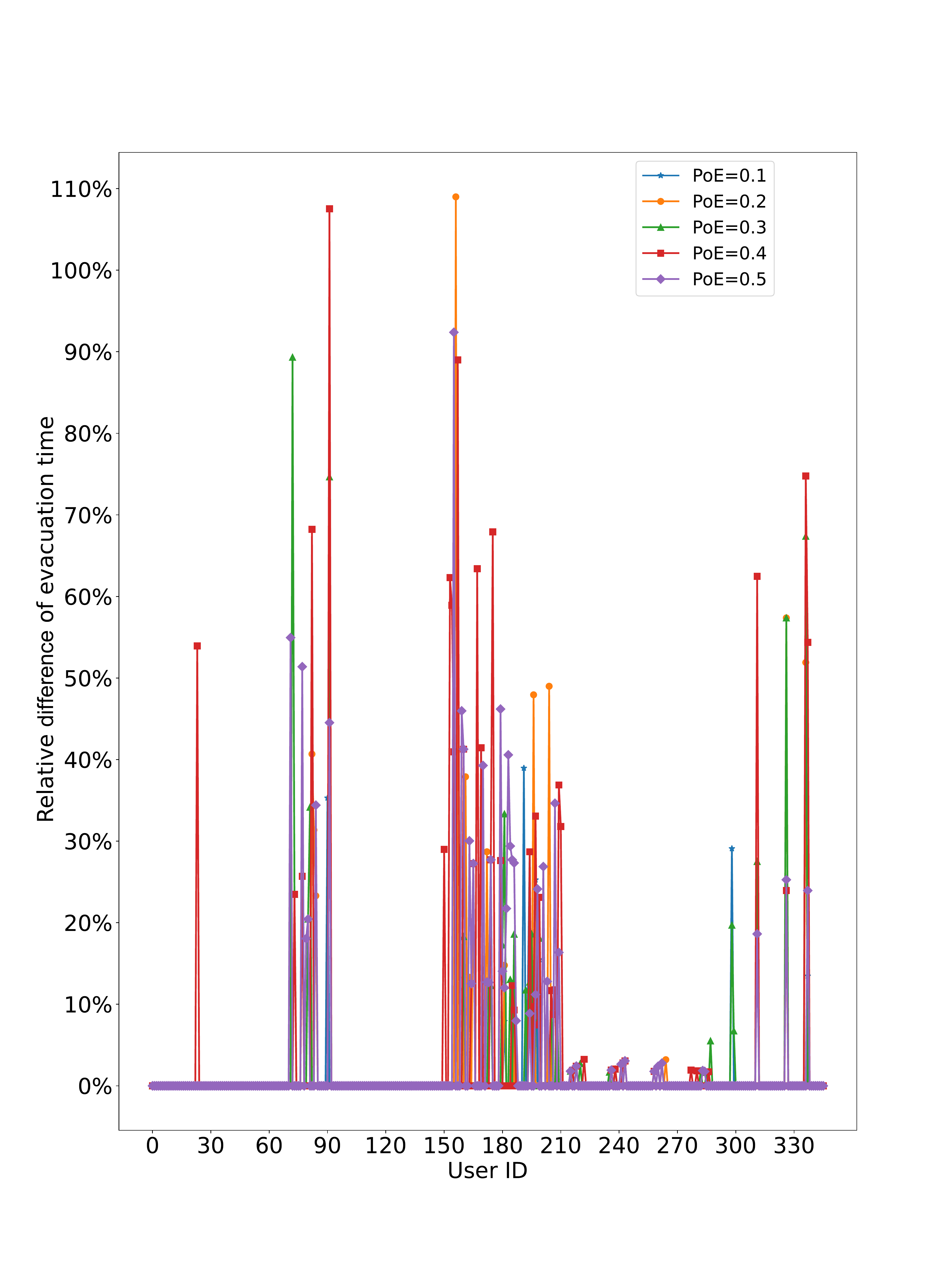} 
  \caption{Evacuation time with different uncertainty probabilities and the relative difference in evacuation time between escaping without behavior uncertainty and with different uncertainty probabilities.} 
  \label{Fig7} 
\end{figure}

Figure \ref{Fig7} plots the relative difference in the evacuation time between escaping according to the navigation directions totally and escaping with a certain uncertainty probability. The relative difference is defined as follows:
\begin{equation}
    \delta(i, PoE)=\frac{\textit{T}(i, PoE)-\textit{T}(i)}{\textit{T}(i)}.
\end{equation}
We can see that the relative differences are positive in all cases, which means failing to escape according to the provided navigation direction impairs the capability of the evacuation scheme.

\begin{figure}[ht]
  \centering 
    \includegraphics[width=3.4in]{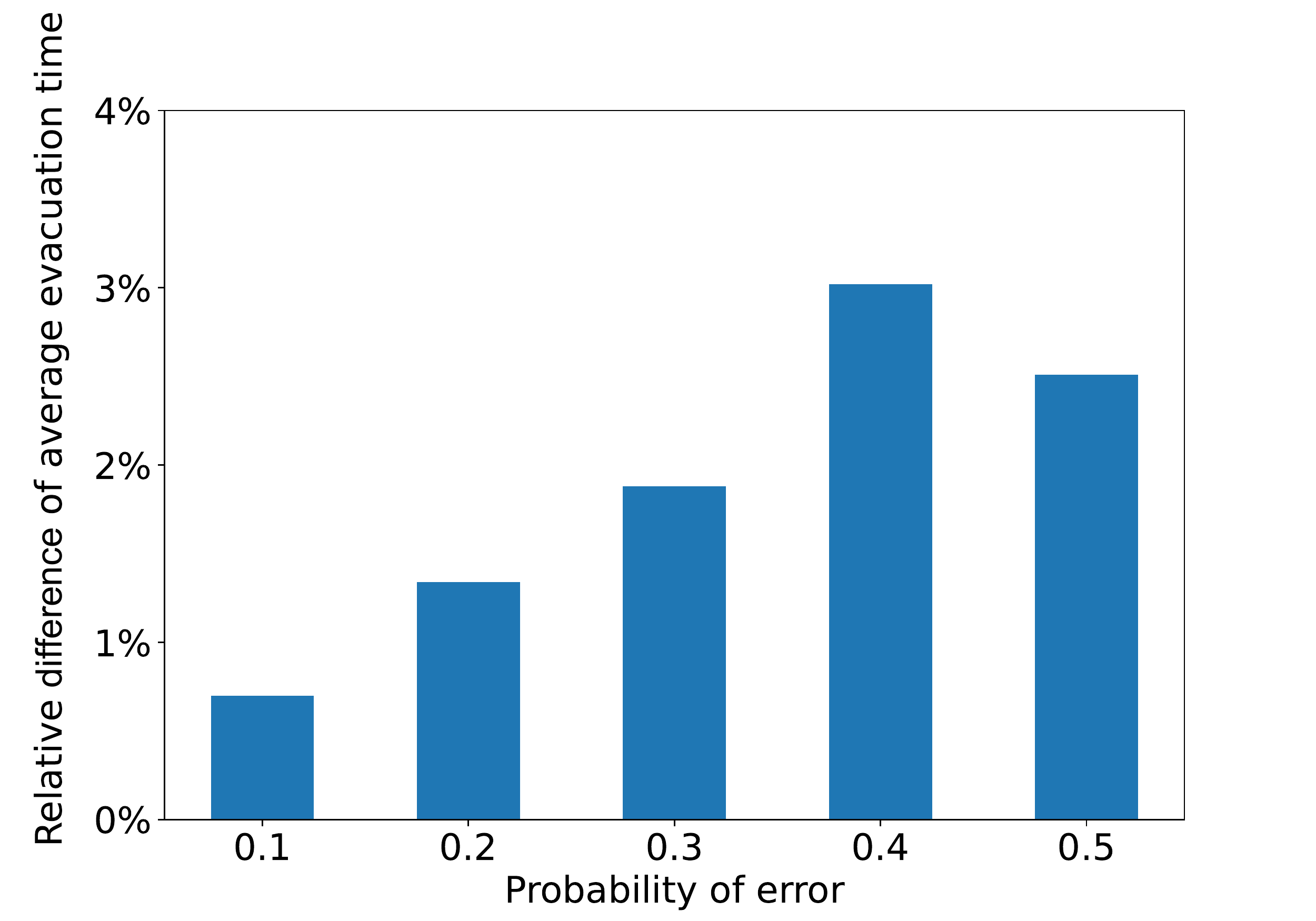} 
  \caption{Relative difference in average evacuation time between escaping without behavior uncertainty and with different uncertainty probabilities.} 
  \label{Fig8} 
\end{figure}

Furthermore, we measure the relative difference in average evacuation time for 346 users, which is calculated as follows:
\begin{equation}
    \delta_{AVG}(PoE)=\frac{\sum_{i=0}^{345}\textit{T}(i, PoE)-\sum_{i=0}^{345}\textit{T}(i)}{\sum_{i=0}^{345}\textit{T}(i)}.
\end{equation}
Figure \ref{Fig8} plots the relative difference in average evacuation time when escaping with different probabilities of behavior uncertainty (i.e., the probabilities of escaping along an error direction). While behavior uncertainty has an effect on evacuation time, it is tiny.  Moreover, we can see that the performance deteriorates with the increase of the uncertainty probability until it achieves 0.4 (i.e., $PoE = 0.4$). That is, in our simulation, a greater probability than 0.4 will not further jeopardize the evacuation performance. Compared with the impact of the delay in navigation service, the impact of behavior uncertainty is relatively slight, which means the lookup table-based method is more resilient to behavior uncertainty than navigation delay.       

\section{Impact of both behavior uncertainty and delay in navigation service}
In this experiment, we evaluate the combined influence of both the navigation delay and behavior uncertainty on evacuation performance measured by evacuation time.

\begin{figure}[ht]
  \centering 
    \includegraphics[width=3.4in]{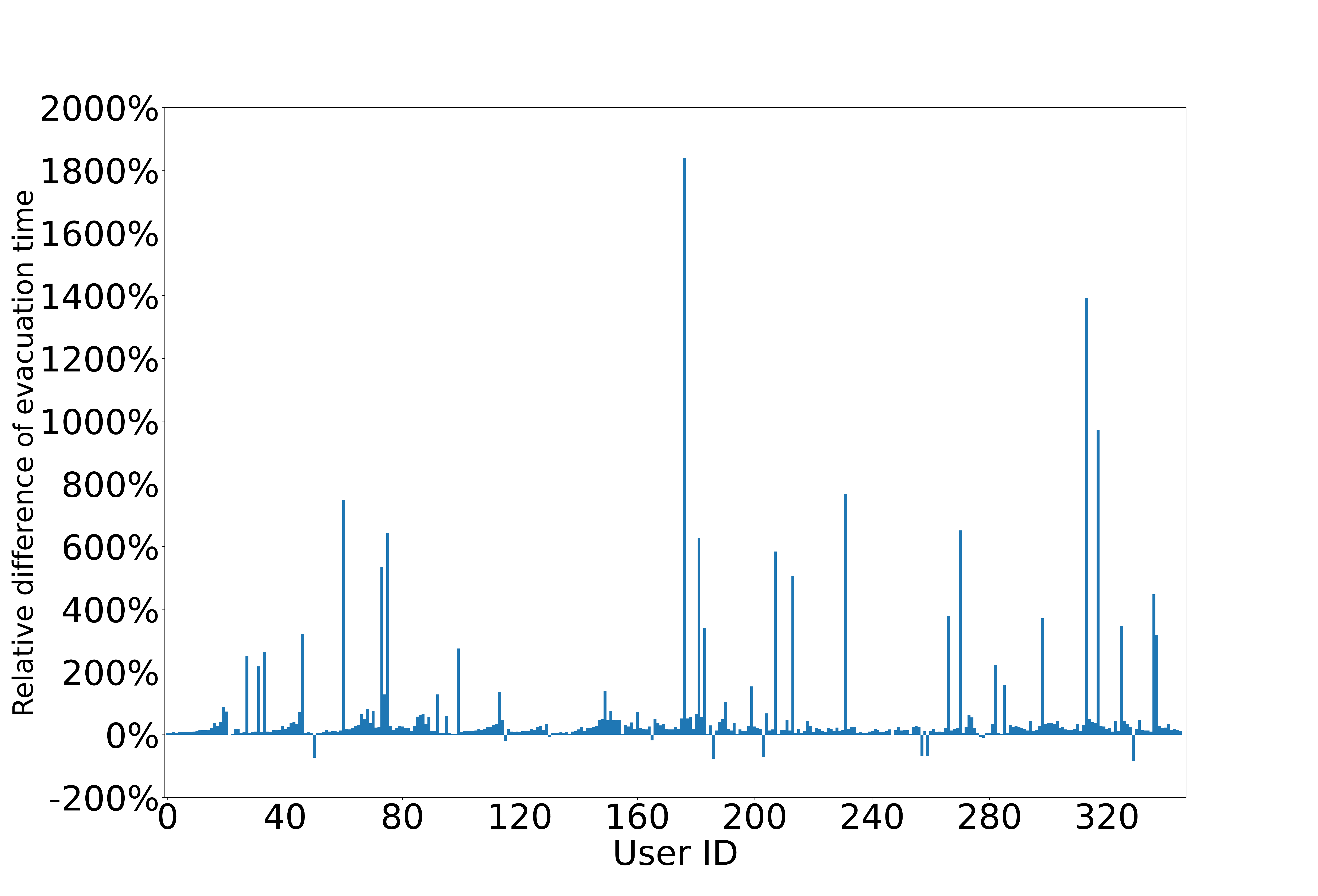} 
  \caption{Evacuation time and the relative difference in evacuation time.} 
  \label{Fig9} 
\end{figure}

The experiment is done in a scenario where the refreshing period is set to 5s, the probability, as well as the delay value, is set as 0.4 and 3, respectively, and the uncertainty probability is fixed at 0.4. Figure \ref{Fig9} plots the relative difference in the evacuation time taken for all users to arrive at the specified exit. We observe that there is a dramatic increment in the evacuation time when we take the above-mentioned realistic features into consideration. Therefore, from a practical perspective, it is of significant value to consider the behavior uncertainty and the navigation delay to assist users with a successful escape while doing path planning for passengers on damaged ships.  

\section{Conclusions and Further Research}

In passenger ships and other large vehicles and aircraft, reliable emergency evacuation is required for ensuring passenger trust in the means of travel, and for their safety and wellbeing.  Thus over the last decade
substantial research has been conducted in designing technology-assisted means to provide
passengers with the best advice regarding evacuation procedures during emergencies
\cite{dimakis2010distributed} and many of the proposed approaches use some form of optimization \cite{gelenbe2013future}.

However during emergencies, especially if the vessel is damaged, it becomes very challenging 
both for the underlying ICT (Information and Telecommunication Technologies) and for the passengers and staff to implement and follow instructions that are based on prior optimization and established routines. In addition, the ICT infrastructure may also be damaged and disconnected, and the recommended evacuation routes and the communication network may be congested. Thus the need to support user safety and evacuation in such complex and dynamic environments with many rapidly changing variables
can also overwhelm the ICT-based emergency management system itself \cite{filippoupolitis2010adaptive}.
In addition, path planning based on the updated emergency situation may be accompanied by messages which flood the communication system about environmental dynamics, aggravate the congestion, and result in packet losses and longer end-to-end delays in communicating decisions.  

Thus this work is the first study of the influence of such key side effects of emergency evacuation in a practical ship evacuation scenario. Through extensive simulations, this paper analyzes the effects on passenger evacuation performance in ship emergencies, of realistic features including the behavior uncertainty represented with different probabilities, the dynamics of traversal time with change frequencies, and the different delays in the arrival of  navigation instructions to passengers. Furthermore, these simulations use real-world parameters from the real passenger cruise ship ``Yangtze Gold 7''  and its passenger evacuation system, and evaluate the effect of delays in the information that reaches the human users during its passenger evacuation. 

Assuming that ongoing situational information is gathered by wireless sensor networks \cite{pan2006emergency,li2009ern,wang2016send}, we have considered the effect of delayed decisions which are computed and forwarded by a central decision ICT system, to all end users as they pass through pre-determined ``nodes'' that guide them towards safe exit points. As these delays increase, we see that the emergency exit times of passengers are often substantially increased. 

In future research we plan to take into account the delaying factors in advance, so as to design novel decentralized emergency navigation systems for guiding passengers to safety, that pre-locate advisory data to passengers in key system intermediate nodes, and also combine centralized decisions together with individual user decision aids with hand-held mobile devices \cite{Kokuti}.

\section*{}
\bibliography{refs} 
\bibliographystyle{IEEEtran} 

\end{document}